\documentclass[12pt]{iopart}
\usepackage{graphicx} 
\usepackage{dcolumn} 
\usepackage{bm} 
\usepackage{epsfig}
\usepackage{comment}
\usepackage[monochrome]{color}
\begin{document}

\title{Charge States of Ions, and Mechanisms of Charge-Ordering Transitions}

\author{W. E. Pickett}
\address{Department of Physics, University of California, Davis, CA 95616, USA}

\author{Y. Quan}
\address{Department of Physics, University of California, Davis, CA 95616, USA}

\author{V. Pardo}
\address{
Departamento de F\'{\i}sica Aplicada, Universidade
de Santiago de Compostela, E-15782 Santiago de Compostela,
Spain
}

\date{\today}

\begin{abstract}
To gain insight into the mechanism of charge-ordering transitions, which conventionally are pictured
as a disproportionation of an ion M as 2M$^{n+}$ $\rightarrow$ M$^{(n+1)+}$ +
M$^{(n-1)+}$, we (1) review and reconsider the charge state (or oxidation number) picture
itself, (2) introduce new results for the putative charge ordering compound AgNiO$_2$
and the dual charge state insulator AgO, and 
(3) analyze cationic occupations of actual (not formal) charge, and work
to reconcile the conundrums that arise. We establish that several of the clearest
cases of charge ordering transitions involve no disproportion (no charge transfer
between the cations, hence no charge transfer), and that the experimental 
data used to support charge ordering
can be accounted for within density functional based calculations that contain no
charge transfer between cations. We propose that the charge state picture 
retains meaning and importance, at least inn many cases,
if one focuses on Wannier functions rather than atomic
orbitals. The challenge of modeling charge ordering transitions with model Hamiltonians
is discussed.
\end{abstract}

\maketitle

\section{Introduction to the Issues}

As the momentous transition between conducting and insulating states by making a
simple change (temperature, pressure, electron concentration, disorder), the 
metal-insulator transition (MIT)
has assumed extraordinary importance in condensed matter physics research and
plays a role in actual and anticipated applications. Established
mechanisms include (1) {\it inter}electronic repulsion of electrons at low density
(Wigner transition), (2) {\it intra}atomic repulsion and resulting correlation (Mott transition),
(3) disorder leading to incoherence and localization (Anderson transition), (4) coupling to lattice,
thereby opening new band gaps (Peierls transition), and 
(5) {\it inter}atomic repulsion leading to charge
ordering (``Verwey transition?"). The latter mechanism is distinguished by a breaking of
symmetry on a sublattice of cations having partially filled, correlated (viz. $3d$) 
electronic shells.

The mechanism for the Verwey MIT in magnetite Fe$_3$O$_4$ 
from the time of its discovery\cite{verwey1939}  was speculated
to involve ordering to two charge states Fe$^{3+}$ and
Fe$^{2+}$ on one Fe sublattice, making it the earliest example of a charge order
transition, one that evidently would explain the MIT. A related charge order mechanism
arises from disproportionation occurring at, and driving, the transition, which might also be
envisioned as the mechanism in magnetite: 2Fe$^{2.5+}$$\rightarrow$ Fe$^{2+}$ + Fe$^{3+}$.
The former is a disorder-order (charge disorder - charge order) transition in which
the ``symmetric'' metallic phase contains two space- and time-fluctuating charge
states, the second
is a disproportionation-driven transition in which every site 
in the metallic phase is equivalent 
even on a short time scale\cite{explain}  (rather than only in time-averaged diffraction). 

In this paper we extend our earlier work\cite{COprl} on analyzing the connection between putative 
charge order driven MITs and the cation charge states that are involved. This study
brings the concept, and the specification, of {\it charge state} or {\it (formal) oxidation
state} or {\it formal valence} to the fore.  We noted that
the physical cation charge -- the $3d$ occupation -- can be identified from electronic
structure calculations, by giving up the usual approach of integrating the density over
some volume and instead simply looking at the cation charge density in the vicinity of
the peak in the radial density $4\pi r^2 \rho(r)$ where only $3d$ density resides. 
Differences in $3d$ occupation, which are our current interest,
are particularly easy to identify and quantify. Several examples of differing
charge states were shown to contain equal $3d$ occupations; actual charge is effectively
divorced from ``charge state'' in some of those examples. 

In Sec. II the original chemist's concept of ``oxidation number'' [or (formal) 
oxidation state] is discussed
briefly, noting that it metamorphosed into a more physical picture of ``charge state''
for materials physicists. A brief poor man's description of what is used to specify the charge state
of an ion is provided in Sec. III.  Some theoretical aspects, relating both to charge order transitions
and to oxidation states, are discussed in Sec. IV.  In Sec. V we begin by recounting 
one illustrative example of
competing charge states in a compound, and proceed to review and extend slightly our 
previous examples of charge ordering materials where it was established that no $3d$
occupation difference between differing charge states exists.
We proceed to an unconventional case -- metallic with second order transition -- ``charge ordering''
compound AgNiO$_2$ in Sec. VI,
and in Sec. VII look into the charge states of Ag in stable, untransforming, insulating AgO.
Our discussion in Sec. VIII pulls together some of our thoughts on understanding and
modeling structural and electronic transitions in the types of systems we have discussed.


\section{Oxidation Number}
Chemists introduced the {\it oxidation number} (sometimes, oxidation state) of an
atom in a molecule almost entirely on the basis of structure and electronegativity.
With its high electronegativity, oxygen has oxidation number -2 except in 
very unusual geometries, 
and halides are -1, in both cases filling a shell. Electropositive ions 
(alkalis, alkaline earths, rare earths) in oxides and halides donate the electrons in
their outer $s$ and $p$ shells, giving obvious positive oxidation numbers
denoted early on and occasionally still by Roman numerals: 
viz. Li$^I$, Ca$^{II}$, La$^{III}$. Most of the complications
arise when, after nature has completed its bonding, an open $d$ shell remains.
For example, vanadium oxides may have V$^{II}$, V$^{III}$, V$^{IV}$, and V$^V$,
and combinations thereof.

The chemistry literature\cite{chem1,chem2,jansen,philipball} contains numerous 
disclaimers that these oxidation states
``have nothing to do with actual charge" and some strive to maintain distinctions
between the different terms in covalent molecules and molecular complexes. 
In the current materials physics literature there is
no discernible distinction. Oxidation states are assigned from the structure
(coordination and distance of neighbors of known valence), 
becoming formalized\cite{Brown2002} in
the {\it valence bond sum} that derives a formal valence (or `bond order') 
directly from bondlengths. Sometimes the magnetic
moment is involved, giving different ionic radii for high spin versus low spin ion. 
Nevertheless oxidation numbers (formal valences) have come to be connected with the
number of actively-occupied local orbitals; through size -- higher oxidation state
means fewer electrons and a smaller ion; through symmetry -- a Jahn-Teller (JT) distorted
coordination shell reflects specific orbital occupations; through magnetic moment, which
requires specific unpaired spin-orbitals.

These oxidation numbers smelled a lot like charges to later scientists (especially
material physicists), and they necessarily sum to zero like real charges. Already in 1939,
fourteen years after the naissance of quantum mechanics, Verwey was discussing the
metal-to-insulator (MIT) transition in magnetite (Fe$_3$O$_4$) -- the Verwey
transition at 120 K -- in terms of charge
ordering of Fe$^{2+}$ and Fe$^{3+}$ ions on one of the iron sublattices. Note that
the notation had already begun to shift Fe(II)$\rightarrow$ Fe$^{2+}$ etc.  The meaning was
clearly {\it physical charges} as opposed to puerile oxidation states: to account for the
transition from itinerant conductor to charge-localized insulator, actual charges
were ordering, getting stuck to ions, becoming localized, and so forth -- that seemed to be the
problem to solve.

\section{Assigning Charge States of $d$ Ions}
Ionic radii and high/low spin designations differ with coordination,\cite{shannon} due 
in large part to the
difference in crystal field.
Unless noted, we use language applying to octahedral coordination for our comments.

{\it Size.} The ionic radius characterization of charge states has organized a
great deal of structural information. The organization of all the
structural data culminated in the table of {\it Shannon ionic radii}
compiled by Shannon and Prewitt that
have provided the standard for more than four decades.\cite{shannon}
For $d$ ions in 6-fold coordination, the
average cation-oxygen distance together with the O$^{2-}$ radius of $\sim$1.40-1.42\AA~
specifies the ionic radius, the Shannon values being the best fit for a given presumed charge
state to a large set of compounds. Note that the ionic radius contains no input
from the actual size of the electronic cloud of the ion. Also, ionic radii differ for different
coordinations, whereas the charge density is virtually invariant.
The charge density is not used, as it is never measured with sufficient precision
to be useful in this context (and the chemists insist it is irrelevant anyway).

{\it Symmetry}. Distortions from symmetric coordination -- Jahn-Teller distortions --
are assigned to occupation of
specific, directional orbitals. Such assignments have been borne out by
electronic structure calculations in a great number of cases. The La$_2$VCuO$_6$
example will be discussed in Sec. V, where it will be seen that spin-orbital occupations
can be confusing.

{\it Magnetic moment.} A Curie-Weiss moment, or ordered moment, reflects ``unpaired''
occupation, {\it i.e.} distinct orbitals are occupied by one spin only. Cations
of a given charge state can sometimes occur in both high spin and low spin (often zero)
configurations. Interestingly, the two spin states that arise in several ions have
been assigned to different ionic radii,
with the high spin configuration typically 0.04~\AA~larger than low spin of a
given charge state.\cite{shannon}
Spin polarization actually changes the
$3d$ occupation and the radial density (real ionic size)  by a negligible
amount. Spin polarization does however create a spin splitting so that up  and down spin
electrons hybridize differently with the O ions, and there is an interorbital Hund's
rule energy.

\section{Theoretical Notions}
\subsection{Modeling of charge order transitions}
Tight binding modeling of charge ordering transitions typically invokes,
beyond the hopping term $H_K$,
direct {\it inter}site repulsion between charges on neighboring cations in
addition to {\it intra}site repulsion:
\begin{eqnarray}
H_U + H_{CO} = U\sum_j n_{j,s} n_{j,-s} + \sum_{<ij>,ss'} V_{ij} n_{i,s} n_{j,s'},
\end{eqnarray}
where $U$ is the Hubbard on-site repulsion and $V_{ij}$ is an intersite repulsive energy, 
where $<ij>$ indicates neighboring sites and $s, s'$ are spin indices. The second
term alone, which is pictured as the driving force, would minimize the energy by
forcing as much charge as possible onto as few distant sites as possible. The Hubbard term
provides the necessary, and physical, energy that opposes excessive inhomogeneity.
Pietig {\it et al.}\cite{Pietig1999} and Hellberg\cite{Hellberg2001} have shown, 
for example, that this Hamiltonian alone, on a two-dimensional
cluster with periodic boundary conditions, shows a range of re-entrant (with temperature)
charge order transitions in a certain range of $V/U$ (with $U$ set equal to the 
bandwidth), similar to the popular interpretation of behavior observed in some manganites. 

Other Hamiltonian terms are included in more general models than the single-band case just
mentioned: very often Jahn-Teller distortion energies $H_{JT}$ for multiorbital 
systems, and sometimes magnetic 
Hund's rule energy (on-site spin-spin) $H_{HR}$ as well as 
distance dependent
kinetic hopping $H'_K$. 
This type of extension (without Hund's rule) has been applied by 
Mishra {\it et al.}\cite{Mishra1999} to model behavior observed in manganites.
In a two band model for $e_g$ electrons in perovskite nickelates (which we discuss
below), Lee and coworkers\cite{Lee2011} applied a Hamiltonian $H_K + H_U + H_{HR}$
(notably without $H_V$) and obtained charge order-like phases driven by Fermi
surface nesting.
These nickelates do not show JT distortions as anticipated of an $e_g^1$ system,
prompting Mazin {\it et al.}\cite{Mazin2007} to suggest avoidance of JT distortion
can be accomplished in certain regimes by charge ordering instead.  They suggested
the form of Hamiltonian mentioned above, minus the $H_V$ term, should provide
the relevant energies.
Interestingly also, when performing a thorough mean field study of ordered states in
$A$NiO$_2$ ($A$ = Na, Li, Ag, the latter of which we address in Sec. VI), Uchigaito
{\it et al.} also chose not to use the $H_V$ intersite repulsion term, although
several other interactions were included. Amongst the
many possible phases uncovered for a selection of independent interaction 
strengths in the Hamiltonian
were phases of charge ordered type.\cite{Uchigaito2011}

\subsection{Identification of charge states / oxidation states}
It is commonly stated and generally accepted that dividing the crystalline 
density into atomic contributions is
so subjective as to be useless for specifying a charge state.
This statement contains a lot of truth but is not entirely correct. Creating disjoint
volumes associated with ions is indeed subjective and ambiguous. However, the general question
is more subtle than dividing up space, so it is
worthwhile to digress briefly.
  
\subsubsection{From pseudoatoms to the enatom}
Almost four decades ago Ball demonstrated,\cite{Ball1975} using small displacements of
nuclei, how the charge density of
a collection of atoms can be decomposed uniquely into contributions from each
atom (``pseudoatoms''). In the process he also identified a charge backflow field associated with
each atom when it is displaced slightly. 
Subsequently he qualified\cite{Ball1977} this result, which does indeed apply to any finite system
and to non-polar solids where the pseudoatoms are necessarily neutral.  
Polar solids, however, provide additional
challenges closely related to the 
difficulties of treating polarization in
such solids, where much progress has been made in the past two 
decades.\cite{resta1994,vanderbilt1993} The fact 
that the Born effective charge for a charged ion
is a tensor provides insight: the (Born) effective charge is not a simple scalar, but
depends on the direction of displacement. 
Ball's prescription has so far been applied only to simple
metals,\cite{Enatom2007} where it may be useful for understanding lattice dynamics
and, applying the same prescription to the potential, 
electron-phonon coupling.\cite{Pickett1979} Because the name {\it pseudoatom} has
been used in so many different contexts, the designation {\it enatom} has been suggested
for Ball's pseudoatom.\cite{Enatom2007}

It remains true that trying to divide space into regions in which the charge
density is assigned to a particular atom is a subjective, hence useless enterprise in practice.
Ball's approach differs, by deriving overlapping atomic densities extending perhaps to a 
few shells of neighbors.
This observation conforms well to our finding,\cite{COprl} discussed at more length below, 
that the $d$ occupation often does not vary at all
between different charge states. If a rather small sphere were chosen to specify 
the charge, different charge states would have identical charges (in those regions). 
Where the charge state picture
clearly works -- which is very often -- the electronic spectrum is a central
property: occupied bands corresponding to occupied local
spin-orbitals can be identified, and the sum of the orbital occupations determines the
charge state. These have conventionally been pictured as atomic orbitals, but the
development and use of Wannier functions (WFs) over the past decade provides an
alternative viewpoint that we return to below.

\subsubsection{Calculation of the oxidation state.}
Two proposals for the definition and calculation of ``oxidation states" (both
used this terminology rather than charge states) have appeared recently.
With molecular complexes in mind, Sit {\it et al.} suggested\cite{Sit2010} projecting
the occupied states onto atomic ($3d$, say) orbitals to obtain the spin-orbital occupation
matrix as is done in the LDA+U method, diagonalize this matrix,
and the number of eigenvalues representing full
occupancy provide the number of occupied physical orbitals and hence the oxidation state.
Smaller occupations are ignored.
For the transition metal complexes they provided as examples, the procedure was
relatively clear.  However, possible ambiguities could be imagined: in one
of the examples 0.93
needed to be interpreted as occupied, 0.63 as not occupied. 
More testing of these ideas
should prove instructive.

Another proposal, by Jiang {\it et al.},\cite{rappe} applying specifically to
periodic systems, employed developments in the theory
of polarization in crystals to formulate their definition of oxidation number as the integral
of (viz. average of) the Born effective charge as the ion's sublattice is displaced
adiabatically by one direct lattice vector. The material must remain insulating but
otherwise the result is independent of the path in configuration space. This procedure does not
identify an ion's charge density or necessarily even  entail a true charge, but does
provide the interpretation of the change in polarization as being due to the transport
of a number of unit charges equal to the oxidation state through a displacement
corresponding to the lattice vector.

As in the case of the enatom mentioned earlier, this definition also does not make the oxidation state
a property of the static reference state -- displacement is required. Because the path is
periodic (it can be repeated), any point can be considered as the starting (reference)
point, with the same oxidation state (same integral). Thus the ion retains one fixed
oxidation state along the path in configuration space. Considering that the Born effective
charge is anisotropic, it seems possible that the oxidation state by this definition
in an anisotropic insulator also depends -- as does the Born effective charge -- on the
direction of the direct lattice vector that connects the end points.
Further applications of this proposal will be enlightening, as will the fact that
displacement is required to specify what is normally envisioned as a property of
the static system.

\subsubsection{Oxidation states from Wannier functions}
In the survey in this paper of charge ordering transition involving different
charge states, and hence of the specification of charge states more generally,
we will begin to explore the viability of abandoning atomic orbitals, whose
occupation often does not change at the `charge ordering' transition, in favor
of Wannier orbitals. The basic idea is: calculate atomic-like Wannier functions for the
relevant states and explore whether these orbitals support the charge state
viewpoint. Though ideas along these lines have been implicit in
discussing specific materials, and Wannier functions have been integrated into
the electronic structure calculations in various ways, this change from atomic
to Wannier orbitals becomes transformational
when applied to `charge ordering' transitions. Without any change in actual
($3d$) charge on the open-shell cation, what are the relevant degrees of freedom
to consider? We return to some of these general issues in the Discussion in
Sec. VIII, but we begin by considering a simple and nearly transparent
case illustrating the idea.

In the double perovskite compound
Ba$_2$NaOsO$_6$, osmium is
formally a heptavalent Os$^{7+}$ $5d^1$ ion -- a very high oxidation state. 
Calculating the $5d$ charge in
the ion, though one cannot arrive at a precise value as noted above, corresponds to
occupation by around 4.5 (not 1.0) $5d$ electrons.\cite{KWLWEP2012} 
Cases of large differences between
actual charge and oxidation state are common knowledge in the electronic structure
community. Does this signal the breakdown of the
charge state picture for high valence state ions? It doesn't: 
calculations\cite{KWL2007,whangbo2007}  indicate
one electron in the otherwise unoccupied ``$5d$ bands,'' spin half of course, and
if the $5d$ occupation is not tied to the charge state (which it isn't) then the
formal valence picture survives in robust fashion.  

Analysis shows that the occupied $5d$ orbital ({\it i.e.} the Wannier function) is a strongly 
hybridized mixture of 50\% Os $5d$ state and 50\% O $2p$ contributions distributed around the 
neighboring oxygen ions.
The other four electrons making up the total $5d$ occupation of 4.5
are hybridized into (and thereby vanish into) the 6 atoms $\times$ 3 orbitals = 18 
occupied O ``$2p$ bands.'' Transformation from atomic
orbitals to Wannier orbitals would presumably lead to 18 occupied O-centered orbitals of $p$
symmetry with tails of Os $5d$ states, and {\it one} occupied Os $5d$-derived orbital
with tails composed of O $2p$ orbitals. So although several $5d$ electrons vanish into
the O $2p$ bands, all five ``$5d$ orbitals'' survive by recovering charge from the 
$2p$ orbitals. Understanding that strong metal-oxygen hybridization implies the efficacy
of generalized (Wannier) orbitals has been around for a while; for example, it provided a
simple picture explaining why {\it second} Cu-Cu neighbor coupling along a
chain dominated over near neighbor
coupling in the Cu-O magnetic chain compound Li$_2$CuO$_2$.\cite{Li2CuO2}  

Another example illustrates an apparent connection between
WFs and formal valence. Volja {\it et al.}\cite{Volja2010} 
have calculated $e_g$ WFs for the two charge states in
a charge ordered manganite La$_{1/2}$Ca$_{1/2}$MnO$_3$, 
finding a substantial difference in the extent of
the tails, the ``Mn$^{4+}$'' WF being more localized. 
Concerns about specifying charge states with WFs include (1) a computational issue:
the gauge freedom means WFs are far
from unique, so subjective choices may lead to different conclusions,
and (2) a conceptual issue: the choice of projections -- the orientations,
for example -- themselves are not objective. 
At one extreme are maximally localized WFs\cite{MaxLocal,MaxLocalRMP} that may not be centered
on an ion and may be unsymmetrical, therefore complicating if not destroying any utility
in representing physical orbitals. Atom orbital- and symmetry-projected WFs\cite{Ku2002} 
(``maximally projected'') provide an alternative approach, one that should
provide a more physical representation, and the projection issue may present less of a
non-uniqueness ambiguity. This viewpoint remains to be tested seriously.

\section{$3d$ Transition Metal Charge States:  Illustrative Examples}
In oxides containing two transition metals, one of the first tasks in understanding them
is to categorize
the charge states of the ions. The sum is known from oxygen and the simpler ions, but a
number of competing factors (which are not well understood and are likely to be 
material-specific) determine the balance.
The double perovskite structure is simple and provides our first example.

\subsection{Competing Charge States of La$_2$VCuO$_6$}
\begin{figure}[!ht]
 \centering
 \includegraphics[width=0.4\textwidth]{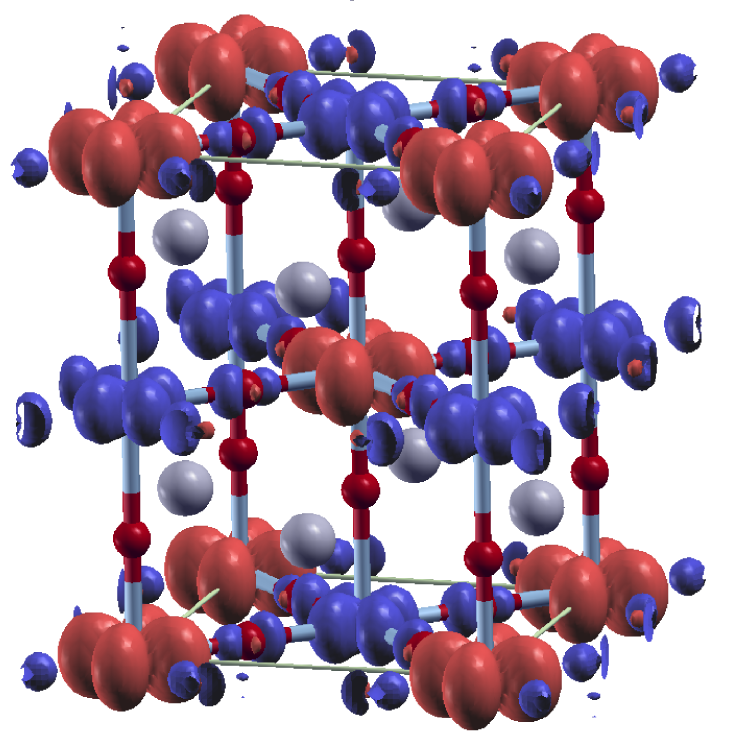}
 \caption{Isosurface plot of the spin density of the JT distorted, spin-compensated
magnetic state of La$_2$VCuO$_6$. The red spin up (say) density in the center of the figure
arises from the V $d_{xy}$ orbital, while the blue spin down density comes from the
unoccupied Cu $d_{x^2-y^2}$ orbital.  The O ions carry a small spin down contribution
due to the $pd\sigma$ (anti)bonding with Cu. La atoms are denoted by the gray spheres.
We note: the $3d$ {\it charge densities}, unlike these spin densities, are not nearly
so anisotropic. For Cu this is understandable because (ideally) only one of ten orbitals
is unoccupied. For V, it is because there is additional, primarily $e_g$ charge occupied
which does not arise in the simple formal picture.
}
\label{VCuSpinDens}
\end{figure}

The computational methods we have used were described previously.\cite{COprl,LAPW,wien}
In our earlier work\cite{vcu} on La$_2$VCuO$_6$, using the LDA+U 
method\cite{anisimov93,Ylvisaker2009} as is necessary for
such oxides, it was found that the two configurations V$^{5+}$Cu$^{1+}$, a standard
closed shell band insulator, and V$^{4+}$Cu$^{2+}$, a spin-compensated ferrimagnetic
correlated insulator
with a very small gap, are practically degenerate in energy after full structural
relaxation. These charge state assignments fit in all respects. Structurally, the mean
V-O distances are 1.84 \AA~and 1.88 \AA~respectively for the two configurations, 
differing by the difference in
the corresponding V ionic radii (0.04 \AA). 
The mean Cu-O distances are 2.02 \AA~and 1.98 \AA, respectively, again
differing by their ionic radii (0.04 \AA~also).  The Jahn-Teller distortion magnitude
for the magnetic configuration (long bond minus short bond) is 0.11 \AA~for both ions,
an unambiguous JT distortion. Magnetically, V and Cu have one electron and one hole,
respectively, as judged by the moment (reduced somewhat by hybridization) and by their
respective filled and empty bands. These two 
configurations are ``poster kids'' for the charge state
picture -- nothing could be clearer. The spin density is pictured in Fig.~\ref{VCuSpinDens}
and supports the charge state description of the spin compensated Mott insulating phase.

\begin{table}[h!]
\caption{Spin-orbital inscribed atomic sphere
occupation numbers from GGA+U calculations for the two
charge states of La$_2$VCuO$_6$. The difference in the Total (last column) for
the two charge states of both V and Cu arise from {\it oxygen tail contributions},
since the $3d$ occupations are identical for the two charge states. Values for
the active Jahn-Teller orbitals are emphasized in boldface.
}
\begin{tabular}{ccccccc}
\hline
Atom               &  $xy$ & $xz$  &  $yz$ &$x^2-y^2$& $3z^2-r^2$ & Total\\
\hline \hline
V$^{4+}:d^1$ up    &{\bf 0.77} & 0.11  & 0.11   &  0.26  & 0.18    &  1.40\\
V$^{4+}:d^1$ dn    &{\bf 0.07} & 0.10  & 0.10   &  0.26  & 0.18    &  0.68\\
V$^{4+}:d^1$ diff  &{\bf 0.70} & 0.01  & 0.01   &  0.00  & 0.00    &{\bf 0.72}\\
V$^{4+}:d^1$ sum   &{\bf 0.84} & 0.20  & 0.20   &  0.51  & 0.36    &  2.11\\
\hline
V$^{5+}:d^0$~~~~   &{\bf 0.30}  & 0.30  & 0.30   & 0.57   & 0.57    &  2.04\\
\hline \hline
Cu$^{2+}:d^9$ up    & 0.94  & 0.92  & 0.92   &  0.95  &{\bf 0.29}    &  4.03\\
Cu$^{2+}:d^9$ dn    & 0.94  & 0.92  & 0.92   &  0.95  &{\bf 0.97}    &  4.70\\
Cu$^{2+}:d^9$ diff  & 0.00  & 0.00  & 0.00   &  0.00  &{\bf -0.68}   &{\bf -0.68}\\
Cu$^{2+}:d^9$ sum   & 1.88  & 1.84  & 1.84   &  1.90  &{\bf 1.26}    &  8.72\\
\hline
Cu$^{+}:d^{10}$~~~~ &1.75   & 1.75  & 1.75   & 1.86   &{\bf 1.86}    &  8.97\\
\hline \hline
\label{TableI}
\end{tabular}
\end{table}

What our earlier work\cite{COprl} also established is that the total $3d$ occupation in both charge
states of V, and also of Cu, are {\it identical} (by this we mean 0.5\% difference or less --
no physically meaningful difference). 
It has occasionally been noted in the literature of electronic
structure calculations (including our own)
that atomic charges in different charge states ``differ rather little''
in charge. In this case, and in the few others that we have checked so far (see below), the 
difference is negligible. 

In Table~\ref{TableI} we display the spin-orbital occupation 
numbers for V and Cu in both charge states
to illustrate some complexities that arise. They are not troublesome in
this straightforward system, but are instructive in that they differ considerably from
what the ideal picture would suggest. These values are $3d$ spin-orbital occupations inside the LAPW
spheres, which contain tails of neighboring oxygen ions and do not include the tails; all
of these tails are in fact ill defined.  The ``total" charge difference
in the sphere
is typical of what is mentioned in the literature: the differences arise from tails of O $2p$
orbitals originating at different distances. For the closed shell, band insulator state
with cubic site symmetry, the $e_g$ occupations for Cu are 6\% larger than for $t_{2g}$, a
difference allowed by cubic symmetry and revealing a density that is mildly deformed
from spherical.  For V however, this difference is nearly a factor of two, reflecting
the bonding of $e_g$ states with O $p\sigma$ states, much of which becomes
hidden in the lower
part of the O $2p$ bands and has little consequence unless considering the actual charge.

The magnetic, JT distorted configurations are more enlightening. Although the charges
are distributed over spin-orbitals in somewhat unexpected  ways as reflected in
Table~\ref{TableI}, the magnetic moment
arises, as the textbook picture would suggest, 
entirely from the JT-active orbitals, $d_{xy}$ in V and $d_{x^2-y^2}$ in Cu.
The moment on both V and Cu is reduced by hybridization with $2p$ states: the majority charge is less
that the formal charge picture suggests, the minority charge is greater, and each
moment has magnitude about 30\% smaller than the ideal spin-half value of 1 $\mu_B$. 

The charge state designation means many things, but it does
not mean at all what the name seems to imply: atomic charge (and spin).  A ``charge state'' 
designation that
does not specify a specific charge seems to classify it as an
oxymoron, yet charge state very often is an essential concept in 
conveying the character of the
ion and its environs (``which
local spin-orbitals are occupied''). Using the term ``oxidation state'' instead of ``charge state''
obscures the issue rather than clarifying it.
It is this conundrum that we will expand on in the following sections. Note at this point that
we have tried not to say  `what {\it atomic orbitals} are occupied' but instead what
local orbitals are occupied.  We provide most of our
discussion that follows, with one exception, in the context of 
materials with ``charge ordering transitions.''

\subsection{Rare earth Nickelates}
The structural and metal-insulator transition\cite{Garcia1992,Torrance1992} in the 
rare earth nickelates $R$NiO$_3$ has
been discussed for two decades as a charge order transition, with experimental data
being analyzed in considerable detail for YNiO$_3$ and several others in terms of a {\it fractional} 
charge transfer 2Ni$^{3+}$$\rightarrow$
Ni$^{3+\delta}$ + Ni$^{3-\delta}$, with the latest 
analysis\cite{Staub2002} pointing to
$\delta~\approx$~0.3 for YNiO$_3$.  The high temperature diffraction-determined structure
has a single Ni site in the distorted GdFeO$_3$-type 
cell containing four formula units with a single Ni site, while
two separate sites Ni1 and Ni2 appear in the low temperature, symmetry broken 
(charge ordered) phase. The 
Ni-O octahedra remain nearly regular (i.e. without JT distortion), with the Ni1-O and Ni2-O mean
distances being 2.02 \AA~and 1.92 \AA, respectively. 
A simple breathing mode mediates the transition. 
The large Ni-O separation differences are consistent
with the Ni$^{2+}$ and Ni$^{4+}$ charge state picture, except for $\delta$=1
rather than $\delta$=0.3.  
There are, however, several questions
about the current interpretation that we now discuss.

\begin{figure}[!ht]
 \centering
 \includegraphics[width=0.5\textwidth]{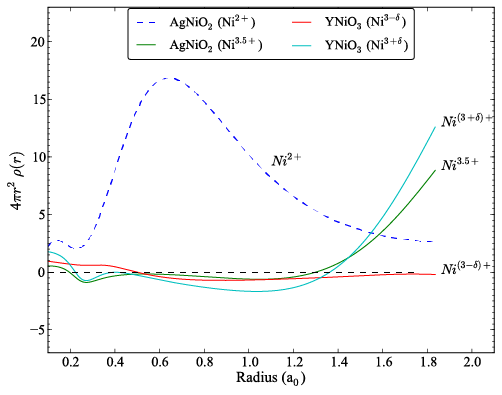}
 \caption{Plot (dashed line) of the radial density of the so-called ``Ni$^{2+}$'' 
ion (Ni1) in AgNiO$_2$, together with
the {\it percentage} differences (with this ion as the reference) 
of the other Ni sites noted in the legend. In the region of the peak the differences
are 0.5\%, even between different compounds.
The differences that become large in the tail region illustrate the sensitivity
in this region to the environment of O $2p$ orbital tails. 
}
\label{radialNidensities} 
\end{figure}

Our investigation into mechanisms of ``charge ordering'' transitions was spurred by
noticing\cite{COprl} that the Ni1 and Ni2 sites have identical $3d$ charge, 
being also the same as in
the high symmetry phase with a single Ni site. In Fig.~\ref{radialNidensities} 
a different type of presentation
is provided. A single ``Ni$^{2+}$'' radial density is plotted, that of one of the 
sites in AgNiO$_2$ (see below); the
other Ni ions in this figure, even those with different charge states, are indistinguishable
on this scale in the region of the $3d$ peak. With this ``Ni$^{2+}$'' site as the reference,
the relative differences (expressed in percent)
 for ``Ni$^{(3-\delta)+}$'' (Ni1 in YNiO$_2$), 
``Ni$^{(3+\delta)+}$'' (Ni2 in YNiO$_3$),
and ``Ni$^{3.5+}$'' (Ni1 in AgNiO$_2$) are plotted. In the region of the peak in density
$r$=0.6$a_0$, the differences are of the order of one-half percent. The shapes of the
density (viz. $3d$ orbital) become slightly different around 1$a_0$, and the effects of
oxygen tails at different distances become evident beyond 1.4$a_0$. 
Note that  the ``Ni$^{3-\delta}$'' density is the nearer to the
``Ni$^{2+}$'' density than are the other two.

\begin{table}[h!]
\caption{Eigenvalues of the spin-orbital occupation matrices for the Ni1 (1.35 $\mu_B$)
and Ni2 (0.55 $\mu_B$) sites in ``charge ordered'' YNiO$_3$, from a ferromagnetic
LDA+U calculation.
Values for ``non-filled orbitals'' are emphasized in boldface.
Notice the absence of appreciable Jahn-Teller splitting (difference between
occupations of $x^2-y^2$ and $3z^2-r^2$).
}
\begin{tabular}{ccccccc}
\hline
Eigenvalue               &  $xy$ & $zx$  &  $yz$ &$x^2-y^2$& $3z^2-r^2$ & Total\\
\hline \hline
Ni1 up    & 0.91   &  0.92  & 0.92   &      0.85 & 0.83  &  4.43\\
Ni1 dn    & 0.91   &  0.91  & 0.91  &{\bf 0.19} & {\bf 0.17}  &  3.08\\
\hline
\hline \hline
Ni2 up    & 0.92   &  0.92  & 0.92  &{\bf 0.64} &{\bf 0.58}  &  3.98\\
Ni2 dn    & 0.92   &  0.92  & 0.92   &{\bf 0.36} &{\bf 0.32}   &  3.44\\
\hline \hline
\label{YNiO3occnumbers}
\end{tabular}
\end{table}

Since the NiO$_6$ octahedra are somewhat distorted in this system, it is the eigenvalues of the
occupation matrices for the two spins for (ferromagnetic) YNiO$_3$ that we present in 
Table~\ref{YNiO3occnumbers}. These values can be used to evaluate the oxidation state
prescription of Sit {\it et al.}\cite{Sit2010} Their prescription gives Ni$^{2+}$ and Ni$^{4+}$
oxidation states as long as one identifies 0.83 and 0.85 as `fully occupied' (the other
fully occupied states have eigenvalue 0.91-0.92), and 0.58 and 0.64 as
not fully occupied and thus {\it un}occupied. To make the issue more apparent, it is useful
to think of scaling the maximum occupation for this sphere size, 0.92, up to 1.00. Then
0.84$\rightarrow$0.91 must be ``fully occupied'' and 0.64$\rightarrow$0.70 must be treated
as ``not fully occupied'' to give the mentioned oxidation states. 
The atomic moments are calculated as
1.35 $\mu_B$ and 0.54 $\mu_B$ versus the idealized values of 2 and 0 $\mu_B$ respectively.

\begin{figure}[!ht]
 \centering
 \includegraphics[width=0.75\textwidth,angle=270]{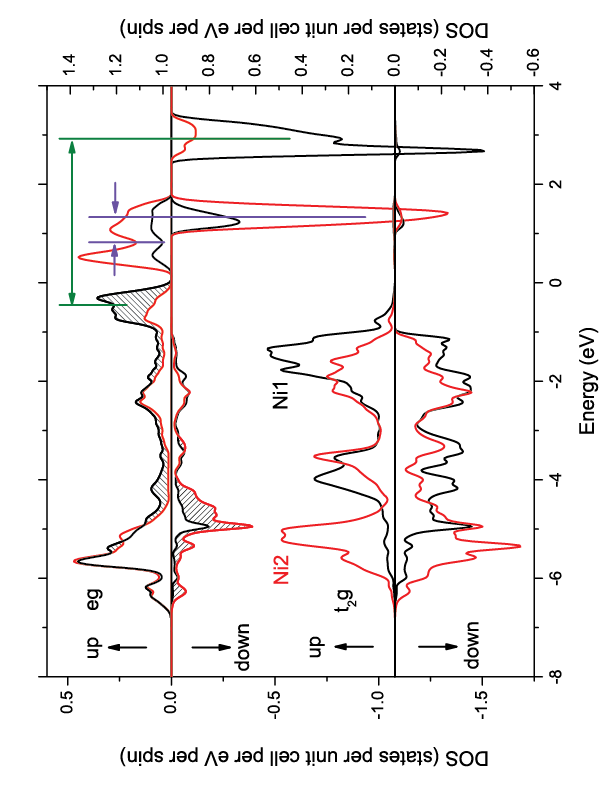}
 \caption{Projected DOS for $e_g$ (above) and $t_{2g}$ (below)
orbitals for both Ni1 and Ni2 sites
in ordered YNiO$_3$. Majority is plotted upward, minority downward. The spectral origin
of the larger Ni1 moment (1.4 versus 0.7 $\mu_B$ for Ni2), which is entirely
from $e_g$ states, is indicated by the hatched regions.
Horizontal arrows indicate exchange splittings, Ni1 in green, Ni2 in purple. 
}
\label{Ni1Ni2PDOS}
\end{figure}

The atomic-orbital projected density of states (PDOS), 
presented earlier,\cite{Mazin2007,COprl} is reproduced in 
Fig.~\ref{Ni1Ni2PDOS}. The larger moment
on Ni1 arises almost equally from more majority states centered at -0.5 eV (an expected
difference, near E$_F$) and less minority states near -5 eV
(an unexpected change in the strongly bound part of the spectrum). 
The amount of {\it unoccupied} $3d$ spectral density does not differ much between Ni1
and Ni2, unlike the formal picture would suggest (a factor of two) but it is 
consistent with our finding of no
difference of $3d$ occupation. The difference lies in the spin and energetic distribution.  
The ``charge state'' picture thus is murky: the ionic radii are consistent with full
disproportionation $\delta$=1  whereas other data (see below) are used to justify $\delta$=0.3
as the best value. The spin-orbital occupation eigenvalues and the PDOS are hardly
definitive.

Very often the contrasting core level energies of different sites are used to identify and
specify charge states, and this has been done for a few nickelates. In our earlier
paper\cite{COprl} we pointed out that these on-site energies are given well by DFT
calculations, and thus are due to the differing environments rather than to any 
difference in local charge.  

\subsubsection{Charge order versus Jahn-Teller distortion}
In charge order transformation phenomenology, or more broadly for the specification of the
oxidation state, the real cation charge -- the
$3d$ occupancy -- is a peripheral, many say irrelevant as mentioned above, 
issue, and our specification
of $3d$ occupation 
reinforces that picture. However, the charge
state {\it language} invites -- in essence, requires -- an interpretation of selected
spin-orbitals to account for differences in the different sites. The ordered state of
$R$NiO$_3$, which contains roughly symmetric NiO$_6$ octahedra of substantially different
size, cannot be understood in terms of an $e_g^1$ ion which should have strong JT tendencies,
hence the conventional ``charge order'' designation. This picture would have the driving
mechanism, the entropy frozen out at the transition, be due to charges hopping amongst
disordered Ni${2+}$ and Ni${4+}$ sites.

Mazin {\it et al.}\cite{Mazin2007} observed that a fully disproportionated 
``high spin $e_g^2$, no spin $e_g^0$''
picture accounts for (1) the lack of JT distortion and (2) the large and small Ni moments.
They confined their attention to the ordered phase (ground state).
The large and small octahedra, expected on the basis of differences in ionic
radii for distinct charge states, does not have any quantitative explanation -- in
any picture -- for ions
of identical charge and radial extent. They further suggested that the Hund's rule
magnetic energy is important in producing these configurations.
Our earlier work established that there is {\it negligible} 
actual $3d$ occupation difference
between Ni1 and Ni2, which seems to leave the search for the true microscopic driving force
between the charge transfer energy, Hund's rule energy, and electron-lattice coupling, 
and the change in kinetic energy
that results from the latter two and modulated by the former.

\subsubsection{Disorder-order scenario}
X-ray absorption spectra for nickelates suggests a very different view of the transition.
Spectral signatures of the two distinct Ni sites are found in the high temperature, symmetric
phase\cite{Piamonteze2005,Medarde2009} as well as beyond the phase transition under
pressure.\cite{Ramos2012} The interpretation must be that the Ni1 and Ni2 sites are already
different {\it above} the transition, but are disordered and fluctuating in both space and time.
The ionic radii are not a good fit for the perovskite structure, and the structure adjusts
by distorting to large and small octahedra. Without structural coherence, carriers can
hop -- the material is conducting although only as a very bad metal. The entropy is due to
fluctuating local breathing modes that freeze in at the transition.

In the ordered phase, the coherent alternation of distortions -- a nonzero amplitude
of the zone boundary breathing mode, to use common language -- opens a gap in the band structure and
the material becomes a (correlated) insulator. The transition can be approached from the
opposite direction: the ordered phase is a small-gap Mott insulator as long as the 
structural distortion is coherent, but loss of the coherence smears the gap and leads to
a very bad metal but conducting phase.  From either viewpoint, the transition is at
the most basic level an order-disorder transition, rendered more complex than most by
the electronic correlation effects. 

The breathing distortion modulates the Ni $3d$ on-site energy, and hence modulates the
charge transfer energy between the O subsystem and each Ni site. JT distortions, on the
other hand, retain the mean Ni-O distance and do not modulate the $3d$ on-site energy
significantly. 


\subsection{Results for Other Charge Order Systems}
In our previous report\cite{COprl} we noted two additional charge-order systems for which
there is no difference in $3d$ occupation of the ``charge ordered'' cations. 

{\it CaFeO$_3$}. This ferrate
is isostructural with the $R$NiO$_3$ class and displays a similar MIT and structural
change. The proposed disproportionation\cite{Takano1991} 
invokes the unusually high oxidation state Fe$^{5+}$
in addition to the charge-balancing Fe$^{3+}$ state. The $3d$ occupations, determined from
the radial charge densities, in the charge ordered
states for the two Fe sites are identical.  It seems plausible that the disorder-order scenario
proposed for $R$NiO$_3$ applies to CaFeO$_3$ as well, though there is much less data to
validate it. Yang and collaborators\cite{Yang2005} noted from their DFT calculations that
no meaningful (i.e. discernible) difference in real Fe $3d$ charge 
(which they identified as around 5.1
electrons) could be obtained, which led them to categorize the ordered state in terms
of ligand holes, $d^5$ + $d^5L^2$ rather than as Fe$^{3+}$ + Fe$^{5+}$. 
The difference in Fe-O distance
changes the hybridization considerably, and the resulting difference in hyperfine field
that they calculated agreed well with the experimental data.  

{\it V$_4$O$_7$.}
Vanadates of many stoichiometries display MITs accompanied by structural transitions,
and the intricacies of the structures complicate identification of the mechanism
underlying the transitions.
The low temperature, ordered phase of V$_4$O$_7$ has been 
characterized\cite{Hodeau1978,Botana2011} in terms of two distinct
V$^{3+}$ sites and two V$^{4+}$ sites. Comparing the radial densities from  
DFT based calculations,\cite{Botana2011} it became clear that all four of 
the V sites have the same $3d$ occupation, as for the various other systems we have
surveyed and discuss in this paper. The difference in deep core levels for the two
Fe sites is 0.9-1.2 eV, similar to the calculated and measured values for the nickelates 
and CaFeO$_3$, and used to substantiate ``charge order."

\section{Charge Order Transition in AgNiO$_2$}
The structural transition at 365 K in this triangular lattice 
compound is not first order but instead continuous, and moreover it is not
a MIT but rather a (semi)metal to (semi)metal transition, so
AgNiO$_2$ belongs to a separate class from the systems discussed above. 
It has nevertheless been characterized 
in some detail as a charge order
system possessing other distinctions. It has been studied also for its unusual low
temperature magnetic behavior and ordering, which are not our primary concern.
The high temperature phase, while conducting, has a resistivity
of 2-3 m$\Omega$ cm, thus classifying it as a very bad metal or, more realistically,
as a semimetal. Decreasing the
temperature from the symmetric phase, the resistivity
{\it increases} slightly but sharply below the transition before leveling off around 40 degrees 
below the transition and again resuming a positive (metallic) temperature derivative.  
This transport behavior is at a second order structural transition, where the freezing
out of lattice fluctuations upon ordering normally leads to a {\it decrease} in resistivity.

The charge state in the symmetric high temperature phase, 
to the extent that one applies in a conductor,
must be Ni$^{3+}$ $e_g^1$, so the triangular lattice of Ni sites  provides a platform 
for frustration of orbital order as well as for potential magnetic order. The transition is
to a $\sqrt{3} \times \sqrt{3}$ increase of the cell caused by oxygen displacement radially
outward (within its basal plane) from one Ni, denoted Ni1, creating a large Ni1O$_6$ 
octahedron and two smaller octahedra
Ni2O$_6$ and Ni3O$_6$. Ni2 and Ni3 sites differ only due to the stacking sequence of
NiO$_2$ slabs, and while this distinction is identifiable in some quantities, the difference
is too small to be of interest here. At low temperature Ni1 has a moment of 
$\sim$1.5$\mu_B$ while the moments on
Ni2 and Ni3 are difficult to quantify, but suggested to be perhaps 0.1$\mu_B$.
The Curie-Weiss moment, both above and for a
wide temperature range below T$_s$, is roughly consistent with {\it either} a moment of
0.7-0.8$\mu_B$ on each Ni (the high T phase value), or 1.5$\mu_B$ on 1/3 of the Ni ions (the
low T phase values). The data and the analysis is not precise enough to distinguish between
these two possibilities, or something intermediate.

\begin{figure}[!ht]
 \centering
 \includegraphics[width=0.75\textwidth]{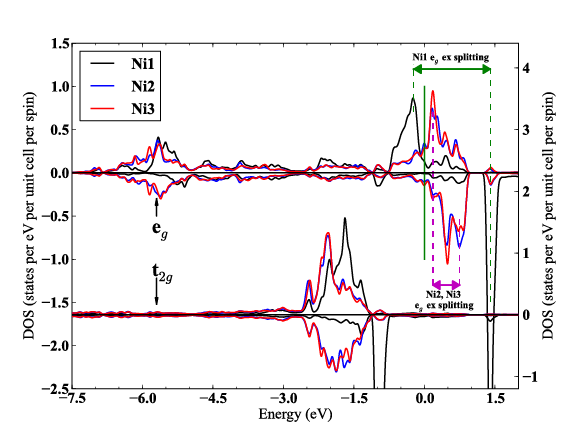}
 \caption{Projected DOS for $e_g$ and $t_{2g}$ orbitals for Ni1, Ni2, and Ni3 sites
in ordered AgNiO$_2$. Majority is plotted upward, minority downward. 
The spectral distributions of Ni2 and Ni3, with calculated moments of 0.28 and 0.27 $\mu_B$,
respectively, show no discernible differences. The Ni1 moment of 1.27 $\mu_B$ is
accompanied by the 2 eV $e_g$ spin splitting, shown at top.}
\label{AgNiO2PDOS}
\end{figure}

Wawrzynska {\it et al.} pointed out\cite{Wawr2007,Wawr2008} 
that DFT calculations indicate that the Ni1 $e_g$ bands
are split $\sim$ 1 eV above and below E$_F$ (Fig. \ref{AgNiO2PDOS}), giving it an S=1 (Ni$^{2+}$)
``Mott insulator'' configuration
(even though no strong correlation corrections need to be used in the DFT calculations), 
with moment reduced by hybridization
to be consistent with the observed value. 
Ni2 and Ni3 are only weakly magnetic, with
their $e_g$ bands crossing E$_F$ (Fig. \ref{AgNiO2PDOS}) 
consistent with an unpolarized ``Ni$^{3.5+}$'' designation
for both ions, necessitated by charge neutrality. 
The O displacement amplitude at low temperature is 0.06~\AA, giving a
change in Ni-O distances due to the distortion of $\sim$0.04~\AA. This amount is perhaps 
reasonable for oxidation states differing by 1/2, but Shannon's  Ni$^{2+}$ - Ni$^{3+}$ difference
is 0.11~\AA, suggesting that a Ni$^{2+}$-Ni$^{3.5+}$ difference should be around 0.15 \AA~
 rather than 0.04 \AA.

As we found for the other compounds which are said to display different charge states, we find
that the $3d$ occupations in ``charge ordered'' AgNiO$_2$, from the radial charge 
densities of the three Ni sites near their peaks, are indistinguishable -- the $3d$ charge
on each is the same. 
More detailed results on AgNiO$_2$, including energetics and magnetic moments during
oxygen displacement, the Fermi surfaces in the paramagnetic state, comparison with available
experimental data, will be published elsewhere.

\section{Charge States in AgO}
Copper with its various charge states has been well studied because of its central
importance in the high temperature superconductors. The isovalent $4d$ sister Ag 
rarely displays correlated behavior in ionic compounds or departs from its monovalent, closed shell
configuration. The behaviour observed in AgO which we now review 
puts it in an unusual borderline class
of materials containing group IB cations.

\begin{figure}[!ht]
 \centering
 \includegraphics[width=0.5\textwidth]{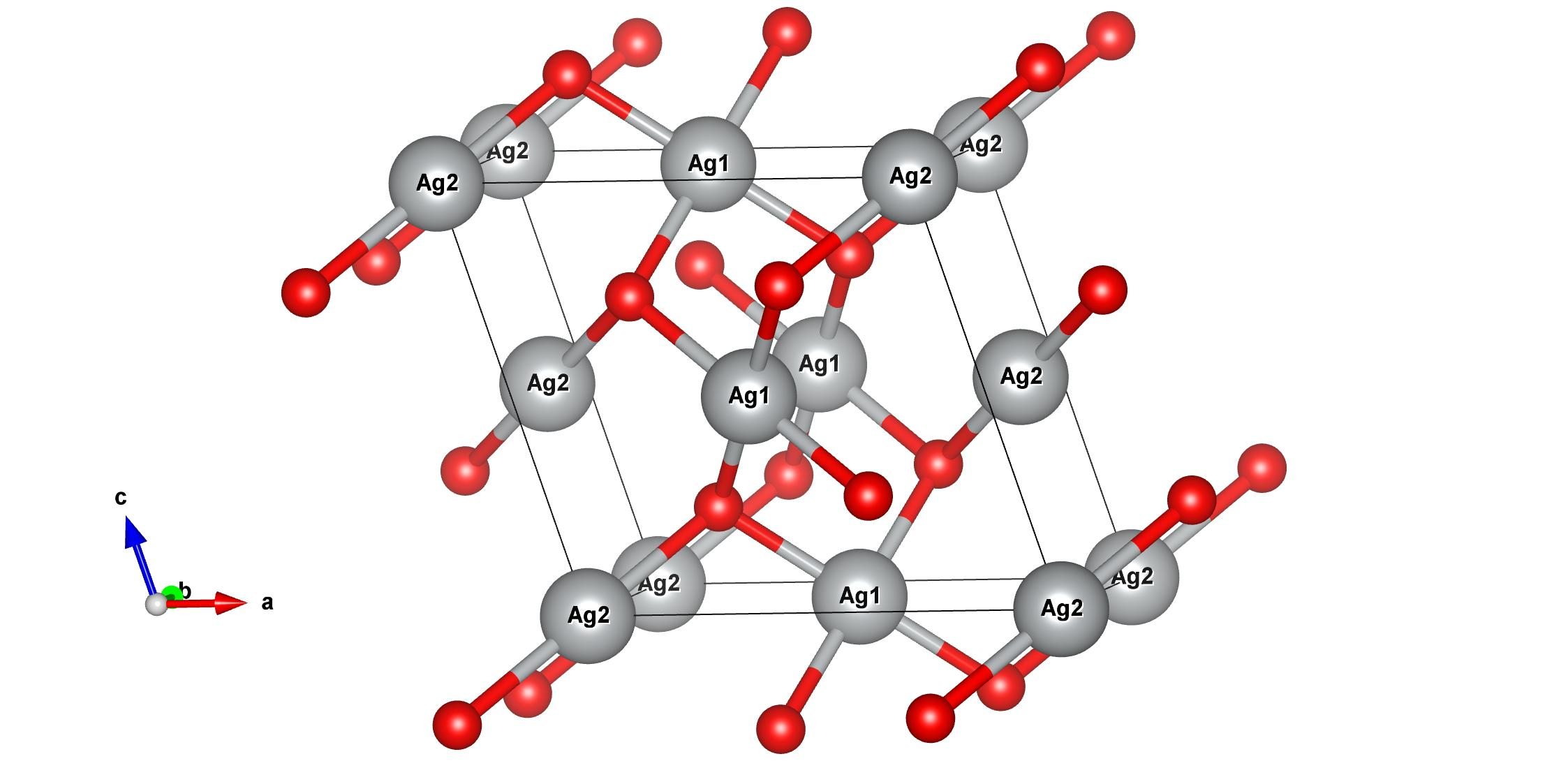}
 \includegraphics[width=0.45\textwidth]{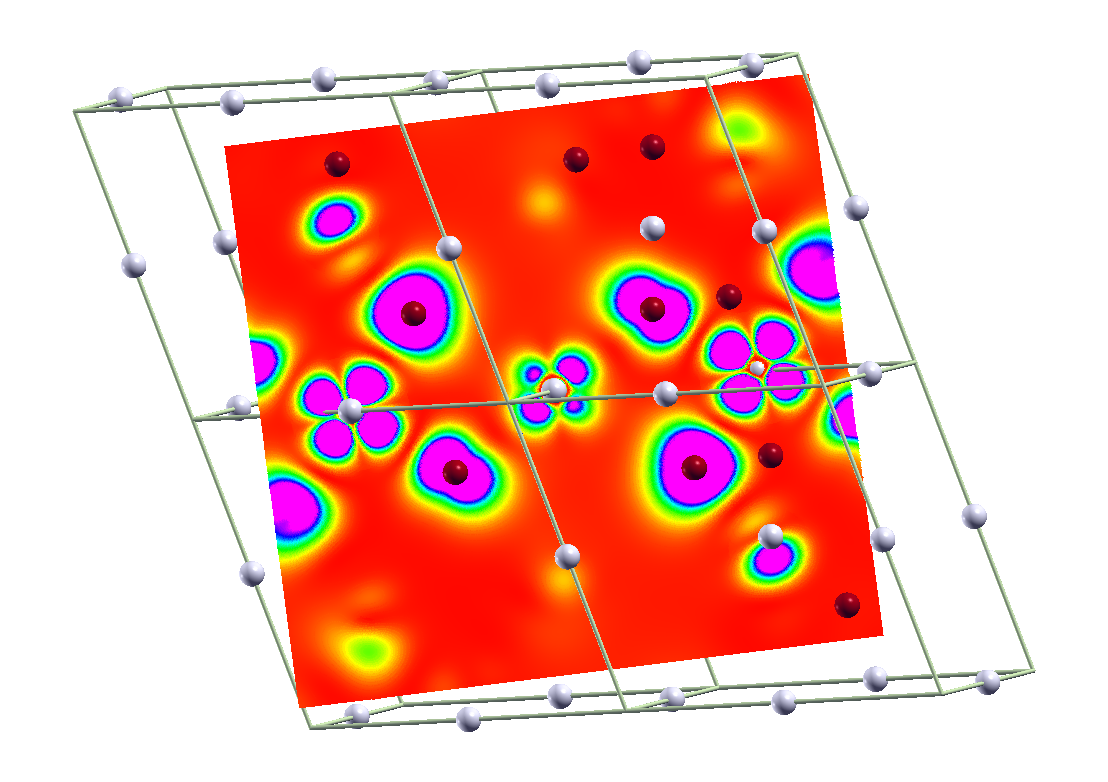}
 \caption{Top panel: crystal structure of AgO, illustrating how the two Ag2O$_4$ squares in the
center are interconnected by the Ag2O$_4$ squares at the top and bottom of the figure.
An O-Ag2-O trimer also connects Ag1O$_4$ squares in a different direction. There is a
single O site, connecting two squares and a trimer.
Bottom panel: contour plot for AgO of the density of unoccupied states in the 0-2 eV region above
E$_F$, through a plane containing Ag1O$_4$ squares on
the left and right sides, connected by an O-Ag2-O trimer that lies in the center
of this plot. White spheres
are Ag atoms, red spheres are oxygen. The cross sections of O give two views of its
strong anisotropy, resulting from directional bonding. The $d_{x^2-y^2}$ ``holes'' on Ag1 in
the square are evident. More surprising the lesser amount, but still very noticeable, 
Ag2 ``holes'' of
$d_{z^2}$ symmetry in Ag1, which is formally a $d^{10}$ closed shell ion.
}
\label{AgOcrystal}
\end{figure}

AgO does not display any charge order or other transition, being insulating and diamagnetic
at all temperatures.
It has however been characterized as displaying both Ag$^{+}$ and Ag$^{3+}$ charge states. The
crystal structure, pictured in Fig.~\ref{AgOcrystal}, is monoclinic P2$_1$/c 
(space group no. 14),\cite{McMillan196028,Jansen1988123}
$a$=5.86\AA, $b$= 3.48\AA, $c$=5.50\AA, $\beta = 107.5$,
with two inequivalent Ag sites, as pictured in Fig.~\ref{AgOcrystal}.
The Ag1 (Wyckoff site $2d$)
sits at the center of a slightly distorted square, Ag2 (site $2a$) forms a linear
O-Ag2-O trimer of length 4.34\AA, and the O ion sits at the Wyckoff $4e$ site.
The average Ag1-O bond length is 2.03\AA, while the Ag2-O separation 2.17\AA, consistent
with a lower charge state and its lower coordination.  Each O (only one crystallographic
site) connects two
AgO$_4$ squares which are almost perpendicular and one O-Ag-O trimer. GGA and hybrid functional
calculations by Allen {\it et al.} were interpreted to support the two charge state
picture: an unoccupied
band is Ag1 derived (one band per Ag1), while all of the Ag2 $4d$ bands are
occupied.\cite{allen2010,allen2011}

\begin{figure}[!ht]
 \centering
 \includegraphics[width=0.50\textwidth, height=0.15\textheight]{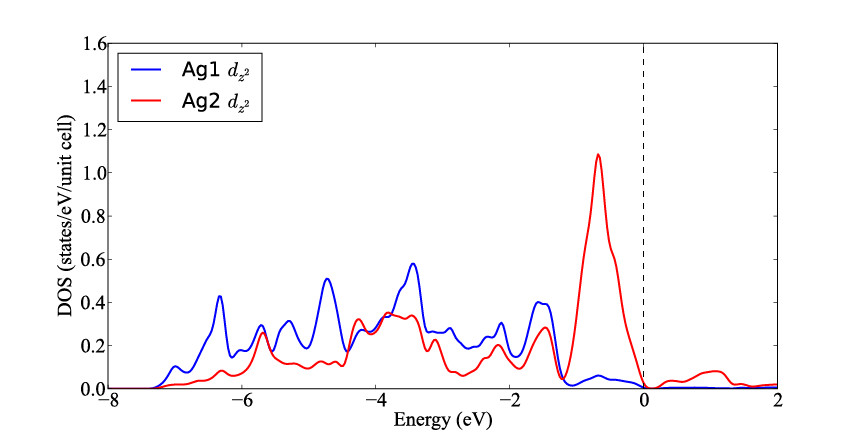}
 \includegraphics[width=0.50\textwidth, height=0.15\textheight]{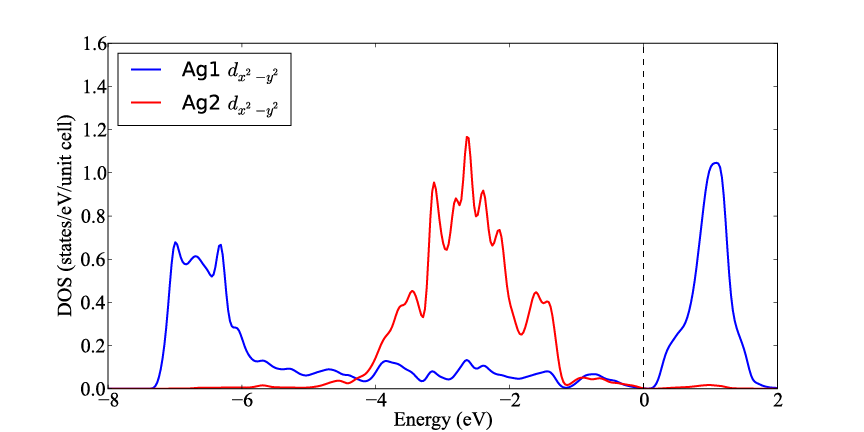}
 \includegraphics[width=0.50\textwidth, height=0.15\textheight]{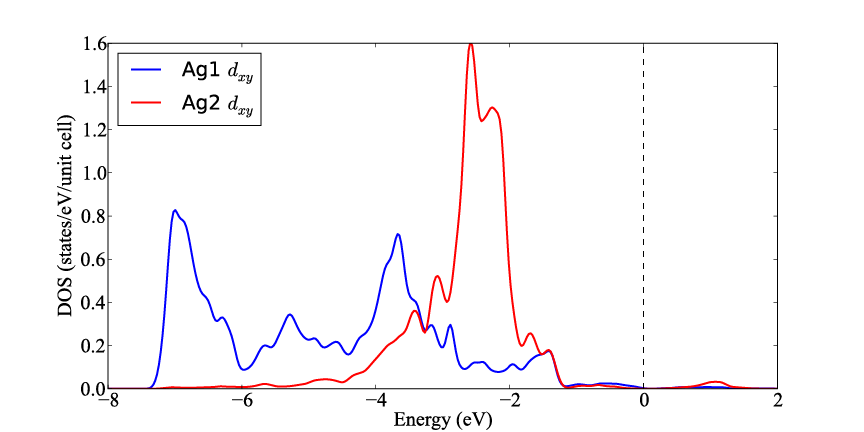}
 \includegraphics[width=0.50\textwidth, height=0.15\textheight]{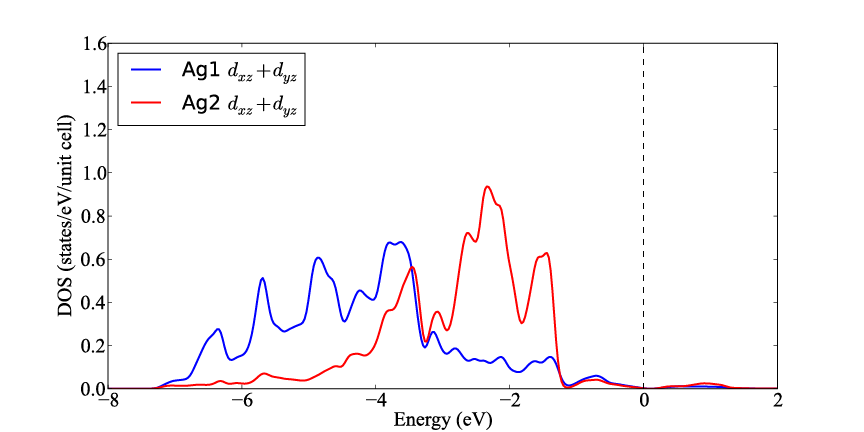}
 \caption{Orbital projected densities of states for Ag1 and Ag2 ions AgO, in the
   natural local coordinate systems of the Ag1O$_4$ square and the O-Ag2-O trimer.
   The Ag1 PDOS is distinguished by the strong bonding/antibonding ligand field
   splitting of nearly 8 eV (second panel from top).
   The Ag2 PDOS has much of its weight in a 2.5 eV wide
   peak centered at -2.5 eV, but is distinguished by the $d_{z^2}$ band lying just
   below the gap (top panel). The oxygen PDOS (not pictured) indicates strong 
   hybridization with both Ag1 and Ag2 throughout the spectrum. 
   }
 \label{dos}
\end{figure}

To check the $4d$ occupation and do some further analysis, we carried out LDA+U (U=5eV,
J=0.68eV) calculations using experimental structure.
The atom-projected DOS (pDOS) is plotted in
Fig.~\ref{dos}.  The Ag1 pDOS dominates in the strongly bound region, from -7.5eV to -3.5eV,
is smaller than that of Ag2 in the -3.5 eV to -1.5 eV, then is nearly negligible (``gapped'')
from -1.5 eV through the bandgap. The unoccupied Ag1-derived band (one band per Ag1 ion)
is however roughly half
O $2p$ character, an occurrence that is sometimes characterized as ``O $2p$ holes.''
This latter picture would then assign the charge configuration Ag$^{2+}$$L$, where $L$ is
a ligand (oxygen) hole, instead of Ag$^{3+}$. Such a viewpoint becomes difficult to sustain,
however, because the seemingly apparent Ag$^+$ charge state requires it to be surrounded
by O$^{2-}$ ions, and there is only the single O ion coordinating with both Ag1 and Ag2.

The Ag2 pDOS is contained primarily in a 4 eV wide peak centered at -2.5eV,
plus the band just below the gap that is half O $2p$ in character. The O $2p$ pDOS is spread
relatively evenly through the range of the $4d$ bands, mixing strongly with both Ag1 and Ag2
except for some Ag2 bands in the -2 eV to -3.5 eV range.

\begin{figure}[!ht]
\centering
 \includegraphics[width=0.75\textwidth]{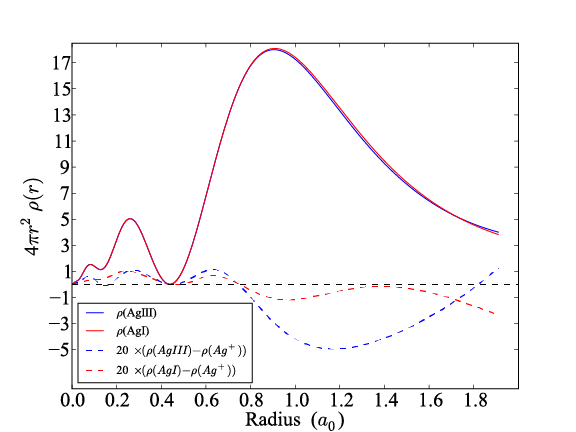}
 \caption{Radial charge density $4\pi r^2 \rho(r)$ of the Ag1 and Ag2 ions in AgO,
and of the Ag$^+$ ion in AgNiO${+2}$ as a reference density(see text). 
Unlike our other examples in ``charge ordering''
materials, there is an appreciable difference of 0.11$e$ in the $4d$ occupation
of Ag1 (Ag$^{3+}$) and Ag2 (Ag$^+$).
}
\label{orbitalproject}
\end{figure}

As in the several cases mentioned above, we find negligible difference in the $4d$ occupation
of actual charge (the radial charge densities are nearly identical near the peak).
This negligible difference in $4d$ charge, as well as the extreme similarity to the
Ag$^{+}$ ion density in AgNiO$_2$, is illustrated in Fig.~\ref{orbitalproject}.

The oxidation state specification of Sit {\it et al.} can be tested here, using the
natural local coordinate system where the orbital occupation matrix is nearly diagonal.
For Ag2, four orbitals have occupation very near 0.80, {\it i.e.} 
nominal full occupation, since
$\sim$20\% of the charge extends out of the LAPW sphere. The other, $d_{z^2}$, is 0.75.
Supposing that 94\% also corresponds to a filled orbital, Ag2 is indeed $d^{10}$, Ag$^+$. For Ag1
in the square, four orbitals have $\sim$0.82 occupation, while the $d_{x^2-y^2}$ orbital
has occupation 0.54, about 68\% full.  Sit {\it et al.}'s prescription would be to
ignore this, obtaining $d^8$, Ag$^{3+}$.

\begin{figure}[!h]
 \centering
 \includegraphics[width=0.35\textwidth]{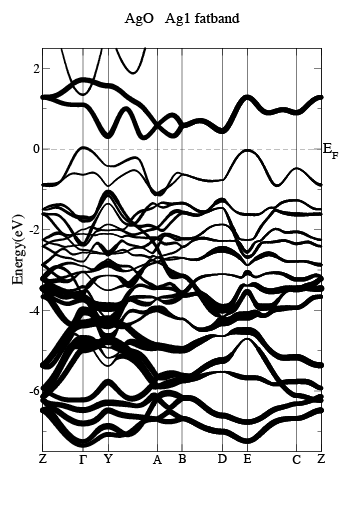}
 \includegraphics[width=0.35\textwidth]{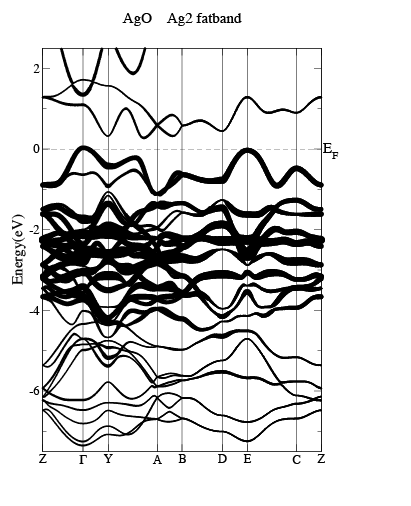}
 \caption{Ag1 (left) and Ag2 (right) fatbands.  The unfilled ``Ag1 band'' (actually
half Ag1, half O, invites the 3+ charge state designation of Ag1. See text for
discussion.}
 \label{fatbands}
\end{figure}

A more illustrative picture is provided by the Ag1- and Ag2-projected fatbands
plotted in Fig.~\ref{fatbands}, which illustrate several characteristics. Much of
the weight of the Ag1 atom lies in the lower regions of the $4d$ bands, which are 
mixed strongly with O $2p$ states. A strong ligand field splitting separates one
orbital -- the $d_{x^2-y^2}$ member that antibonds strongly with the neighboring
O $2p$ orbitals.  
The Ag2 1+ ion in the trimer has most of its weight from
-4 eV to -1 eV, with a just split off antibonding $d_{z^2}-p_z$ combination forming a 1 eV
narrow band below the gap. 

\section{Discussion}
Briefly stated, we have observed that for several cases of ``charge ordering'' in
solids, usually involving a first order metal-insulator transition,
there is no change in cation $3d$
occupation across the transition, hence no real charge ordering.  This lack of any
charge transfer between cations during a ``charge ordering'' transition raises
fundamental questions about the mechanism(s) behind these transitions and the
best way to understand the underlying physics. There has
been some discussion of a similar conundrum in magnetite by Garcia and 
Subias.\cite{Garcia2004} At a more
basic level, this observation raises perhaps more basic questions about the 
meaning and specification of ``charge states'' in solids in addition to the 
underlying mechanism -- just what it is that orders at the transition, where is
the entropy gained or lost?

As we have surveyed these few systems, we have observed  that if Wannier 
functions rather than atomic
orbitals are used as the physical local orbital, the charge state picture may survive as
a useful characterization at least in most of the cases. 
The objectivity and generality of this viewpoint remains to
be tested, and there are cases such as the $R$NiO$_3$ nickelates where the resulting
picture seems less than convincing even with Wannier functions in mind. 
Likely other gray areas will be found as these questions are pursued; fractional
charge ordering is especially problematic.

Returning to another underlying motivation -- the faithful modeling of ``charge ordering''
transitions -- the basic issues have changed as the microscopic behavior begins to be
understood. The initial question we posed was: what are the appropriate terms in a model
Hamiltonian to provide correct modeling of the charge-spin-orbital behavior 
through the transition that
leads to ``charge ordering transitions''?
The answer is that this seems not to be the right question. The
challenge is different: in a Wannier function basis of $3d$ orbitals the most
transparent local orbital basis, with occupations one or zero, is different
on either side of the transition -- the environment is instrumental in characterizing
the charge state. The difference can be seen in several ways, but
most clearly by the observation that the $3d$ occupation would differ 
(by order unity, the difference in occupations)
if they were not considerably  different. A physical local orbital description
on each side of the transition is simple only if substantially different local orbitals
are used on either side. A basis sufficient to enable both representations
is the set of all $3d$ and $2p$ orbitals
(other orbitals are out of the picture);
a single simple $e_g$ or t$_{2g}$ basis lacks the necessary flexibility. The issue
is not what terms are in the Hamiltonian so much as that a good minimal basis 
on one side of the transition is a poor, insufficient one on the other.

Einstein is said to have stated: ``a theory (or model) should be as simple as 
possible, but no simpler.''
The requisite basis set, until shown differently, is (1) the set of relevant atomic $d$ states on the open
shell cation ($e_g$ or $t_{2g}$ when crystal field splitting is large, for example)
{\it and} (2) the active $2p$ atomic orbitals on the ligand (oxygen). These O functions
are necessary because the O participation in a Wannier function changes across
the transition, and this is a crucial degree of freedom. 
The on-site
repulsion $U$ is relevant due to its role in keeping the $d$ occupation
fixed, but perhaps it can be thought of heuristically as a $U\rightarrow\infty$ term as far as the mean field
physics is concerned.  Cation-ligand hopping terms and cation-ligand Coulomb energies,
both distance-dependent, must be important. The on-site energy difference
(``charge transfer'' energy $\varepsilon_d - \varepsilon_p$) of a 
reference state (the high symmetry structure) is
important in determining the character of the Wannier functions, and Hund's rule energy
is a relevant factor.


Setting up and carrying out fruitful modeling is a task for the future.
A fundamental question remains: why does a given charge state -- occupation of a specific
type of Wannier orbitals -- lead to a reasonably well defined ionic radius, rather than a
continuum depending on the environment --
why are ionic radii `quantized?'
More broadly stated: why does the conventional charge state picture function so
usefully, when $3d$ (or $4d$) charge is not involved -- both cation and anion charges are
unchanged?\cite{quantized} 
An understanding of ``charge ordering''
transitions may require an answer to this question.

\section{Acknowledgments} The authors have benefited from discussions with a number
of colleagues, notably P. B. Allen,  D. Khomskii, and P. Kalifah. This work was
supported by National Science Foundation award DMR-1207622-0.


\begin{thebibliography}{5}
\bibitem{explain}The distinction may be one of time scale of electron hopping between
Fe sites, which will be reflected in the magnitude of the resistivity in the conducting
phase.
\bibitem{LAPW} The density functional based linearized augmented plane wave code
{\sc wien2k}\cite{wien} has been used in our
calculations, where we have applied correlation
corrections (LDA+U) as is usually necessary to describe $3d$ oxides. Computational
methods are as described in Ref. \cite{vcu}.
\bibitem{quantized}In truth, there is some spread in a given ion's `ionic radius' in different
compounds, but approximate quantization is observed.
\end{thebibliography}
\end{document}